# Applications of (Wigner-Type) Time-Frequency Distributions to Sonar and Radar Signal Analysis

Guillermo G. Gaunaurd[†]
*Army Research Laboratory, Code AMSRL-SE-RU, 2800 Powder Mill Road, Adelphi, MD 20783-1197*

Hans C. Strifors[‡]
*Defense Research Agency (FOI), P. O. Box 1165, SE-5811 Linköping, Sweden*

## Abstract

Wigner-type distributions have shown their effectiveness in classification problems of sonar and radar. We present an overview of applications where resonance features in echoes scattered from targets insonified or illuminated with short pulses are studied in the joint time-frequency domain. We first show the acoustic case of an elastic shell with a few filler materials submerged in water and insonified by the "clicks" dolphins generate when they are echo-locating the targets to classify them. The second example deals with a ground-penetrating radar used to identify buried land mines. We have compared the performance of various time-frequency distributions as well as a technique combine these with a "fuzzy-cluster" representation of the subsurface target signatures.

## I. INTRODUCTION

The Wigner distribution (WD), which has received considerable attention since its introduction in 1932 [1], is an excellent example of Wigner's well-known statement about "the unreasonable effectiveness of mathematics to describe the physical world." This marriage of the ambiguity function and the Fourier transform is today the foundation of the field of time-frequency (TF) representations of signals. Moreover, it is particularly important in the area of target recognition. Signals backscattered from targets are just more amenable to analysis and interpretation in the TF domain. Of the various ways this can be done, it has been our observation that the WD and its close relatives, such as the pseudo-Wigner distribution (PWD), are excellent candidates to pursue such active (or passive) classification tasks. In this work we show its effectiveness in classification problems of sonar and radar. Resonance features in target echoes are well-known entities to look for to identify unknown targets. These can be studied in either the frequency or the time domains. However, a time-frequency distribution (TFD), which is a two-dimensional function, can reveal the time-varying frequency content of one-dimensional signals. This proper-ty of the TFDs is particularly useful for the extraction of signature features from (transient) target echoes.

---

[†] electronic mail: ggaunaurd@arl.army.mil
[‡] electronic mail: strifors@foi.se

We first show the acoustic case of an elastic shell submerged in water and insonified by the "clicks" dolphins use when echo-locating targets to classify them. We cannot enter the dolphin's brain, but we believe they use a TF approach to carry out their amazing feats. Anyway, we can, in these examples, distinguish between the submerged targets using a PWD. The second example deals with a ground-penetrating radar (GPR) used to identify buried land mines. Much effort has been devoted to this case, theoretical and experimental. We have compared the performance of various TF distributions as well as a technique to combine the PWD or any other time-frequency distribution (TFD) with a "fuzzy-cluster" representation of subsurface target signature.

## II. THEORETICAL BACKGROUND

### 1. TIME-FREQUENCY DISTRIBUTIONS

In a wide variety of applications, bilinear TFDs are a primary signal analysis tool. Each one of the large number of members in the general class (or Cohen class) of bilinear distributions [2,3] that has been proposed can be interpreted as the two-dimensional Fourier transform of a weighted version of the (symmetric) ambiguity function (AF) of the signal. These distributions can all be written as

$$C_f(t,\omega;\Phi) = \frac{1}{2\pi} \int_{-\infty}^{+\infty}\int_{-\infty}^{+\infty} A(\theta,\tau)\Phi(\theta,\tau)\, e^{-i\theta t - i\tau\omega}\, d\theta\, d\tau, \qquad (2.1)$$

where $A(\theta,\tau)$ is the AF of the signal $f(t)$ given by

$$A(\theta,\tau) = \int_{-\infty}^{+\infty} f(t+\frac{\tau}{2}) f^*(t-\frac{\tau}{2}) e^{i\theta t}\, dt, \qquad (2.2)$$

where an asterisk denotes complex conjugation.

The weight function $\Phi(\theta,\tau)$ is called the kernel, and it determines the specific properties of the distribution. Particular choices of the kernel function defines the various TFDs that have been introduced. The kernel of the WD is constant: $\Phi_{WD}(\theta,\tau) = 1$ and the Choi-Williams distribution (CWD) is defined by the exponential kernel: $\Phi_{CWD}(\theta,\tau) = \exp(-\theta^2\tau^2/\sigma)$. The kernel of the WD passes both auto- and cross-components from the AF into the TFD without attenuation. The kernel of the CWD suppresses the cross-components at the expense of some smearing of the auto-components [4], both to a degree that is controlled by the parameter $\sigma$.

The degree of cross-terms interference that to some extent can be controlled in the CWD by the built-in parameter $\sigma$ can also be controlled by introducing a smoothing window function in the WD. The resulting pseudo-Wigner distribution (PWD) is then given by

$$\text{PWD}_f(t,\omega) = \int_{-\infty}^{+\infty} f(t+\frac{\tau}{2}) f^*(t-\frac{\tau}{2}) w_f(\frac{\tau}{2}) w_f^*(-\frac{\tau}{2}) e^{-i\omega\tau} d\tau, \qquad (2.3)$$

where $w_f$ is the window function. A Gaussian window is: $w_f(t) = \exp(-\alpha t^2)$ and $\alpha$ is the window width controlling parameter.

For comparison we also use the CWD, which can be written in the form [2]:

$$\text{CWD}_f(t,\omega) = \int_{-\infty}^{+\infty}\int_{-\infty}^{+\infty} \frac{1}{\sqrt{4\pi\tau^2/\sigma}} e^{-\frac{(u-t)^2}{4\tau^2/\sigma}} f(u+\frac{\tau}{2}) f(u-\frac{\tau}{2}) e^{-i\omega\tau} du\, d\tau. \qquad (2.4)$$

The smaller the value of the parameter $\sigma$ is chosen, the larger the amount of smoothing and reduction of cross-terms interference. For a large value of $\sigma$ the CWD approaches the WD. It should be noted that cross-terms interference between different portions of a noise-free signal backscattered from a target generates desirable signature features in the TFD, while cross-terms interference associated with added noise gives rise to undesirable "artifacts" that tend to obscure the target's signature.

In the calculations of TFDs we will throughout use the complex-valued analytic signal $\tilde{f}(t)$, which is related to the given signal $f(t)$ by:

$$\tilde{f}(t) = f(t) - \frac{i}{\pi}\text{p.v.}\int_{-\infty}^{+\infty} \frac{f(\tau)}{t-\tau} d\tau, \qquad (2.5)$$

where p.v. denotes the Cauchy principal value of the integral and the imaginary part is the Hilbert transform of the real part. The advantage of the analytic signal is that it has a vanishing Fourier transform for negative frequencies [2,3].

## 2. *TARGET IDENTIFICATION USING RESONANCE FEATURES*

The isolation and extraction of resonance features from sonar or radar cross-sections is done because resonances serve to identify remote targets like their fingerprints. The spectral lines of a target and their widths (i.e., the isolated resonances) can be viewed as a macroscopic spectroscopy scheme [5]. A long pressure pulse with fixed carrier frequency $\omega_0$ incident on an acoustic or elastic target will excite it near the frequency $\omega_0$, since long pulses have narrow spectra. If $\omega_0$ is chosen to coincide with a natural resonance of the target, then the target resonates at that frequency. If not, then no resonance feature will appear in the echo. Short pulses that have broad spectra produce transient scattering effects over the frequency band of the incident pulses. This can be used in impulse or ultra-wideband (UWB) sonar / radar. In the case of sonar, if $a$ is some characteristic length of the target, $c$ the sound speed in the medium, and $k = \omega/c$ is the wave number, then the far-field backscattered pressure returned by the target is [5,6],

$$p_{sc}(r,t)\frac{2r}{a} = \frac{1}{2\pi}\int_{-\infty}^{+\infty} G(\omega) f_\infty(\omega) e^{i(\omega t - kr)} d\omega, \tag{2.6}$$

where $G(\omega)$ is the spectrum of whatever pulse $g(t)$ is incident on the target, and $f_\infty(\omega)$ is the (far-field) backscattering "form-function" of the target. This is the connection between the steady-state (i.e., $f_\infty$) and the transient (i.e., $p_{sc}$) solutions. All that is needed to change from the steady-state situation in the frequency domain to the transient situation in the time-domain (Eq. 2.6) is the convolution of the form-function with the pulse spectrum.

In the corresponding electromagnetic (EM) case of radar where a plane EM wave is incident on a target with a characteristic length $a$, the far-field backscattered electric field is given by [7]:

$$E_{sc}(r,t)\frac{2r}{a} = \frac{1}{2\pi}\int_{-\infty}^{+\infty} G(\omega) f_\infty(\omega) e^{i(\omega t - kr)} d\omega, \tag{2.7}$$

where $G(\omega)$ is the spectrum of the electric pulse $g(t)$ incident on the target, and $f_\infty(\omega)$ is the (far-field) backscattering form-function of the target for the considered polarizations. Here, $c$ is the speed of light in free-space.

## III. AN APPLICATION TO ACOUSTICS

### 1. ECHOES GENERATED BY THE DOLPHIN SONAR

Dolphins have an unsurpassed ability to identify underwater objects using their biological sonar [8,9]. We have examined a large collection of sonar pings ("clicks") emitted by trained dolphins when they are attempting to insonify or "echo-locate" certain targets presented to them by the trainers. The database also contains the backscattered echoes returned to the animal. The experiments were conducted at the Navy NRaD (Hawaii) Laboratory. The experimental set-up has been described elsewhere [10]. The dolphin, with his head in a hoop, was echo-locating on a target placed in the water in front of him. The targets were cylindrical, aluminum shells suspended vertically at a distance of about 4.5 m. The shells were filled with various substances (glycerol, beeswax, or salt water), which the animal was supposed to distinguish. The animal communicates to us his consistently correct choices by pushing an underwater paddle whenever the "correct" target was presented to him. This correct target was the one he had been trained to identify by means of standard conditioning techniques and food reinforcements. A recording system captured all the clicks and the returned echoes. We compared recorded data with our predictions for the same conditions, which showed good agreement. We also used inverse scattering techniques that we have developed that allow the identification of the shell characteristics from selected features in the echoes. There are several spectral features, and their time-domain counterparts, that are observable in all responses. We have shown that humans can extract the main geometrical and physical characteristics of the targets (i.e., shell shape, size, material composition of the shell and its filler, etc.) from these features. Hence, since these clues are also available to the dolphin, we would like to believe, that this is the simple procedure the dolphins use to carry out their amazing identification feats. Basic properties of the incident clicks that best bring out the above-mentioned identifying features in the echo are that the pulses are short (i.e., broadband), of suitable frequency content to match the size of the target, and energetic enough (i.e., of high S/N) to overcome the ambient noise levels that may be present.

## 2. *TIME-FREQUENCY DISTRIBUTIONS TO IDENTIFY DOLPHIN INSONIFIED TARGETS*

Figure 1 displays the 3-D plots of the modulus of the PWD as obtained from Eq. (2.3) with a narrow time window. Figure 1, left (or right) plot corresponds to the theoretically predicted (or measured) backscattered echoes from the shell [11]. In this case the shell is filled with glycerol. The figures also show the 2-D projection in the bottom time-frequency plane, as well as the returned waveform and its spectrum on the lateral planes. Using Eq. (2.6), the theoretical prediction of the returned echo is computed for a cylindrical shell of infinite length but with the same radial dimensions and the same material as the shells used in the experiments. Here, the actual dolphin click is made to insonify the shell. The large "mountain" in each PWD plot exhibits great resemblance. That feature is mainly generated by the specular return, and it shows the broad bandwidth of the dolphin's click. The information contained within this frequency band of backscattered echoes (the specularly returned portion of the echoes and the possible subsequent ringing) is sufficient for the dolphin's successful target identification. Figure 2, left (or right) plots display the corresponding measured response when the shell is filled with beeswax (or salt water). Comparing the three measured PWDs of Figs. 1 and 2, a minimally trained observer can distinguish between these three targets. A more detailed analysis of the information about geometry and material composition of the shells and their filler material was given elsewhere [10,11], and does not fit here.

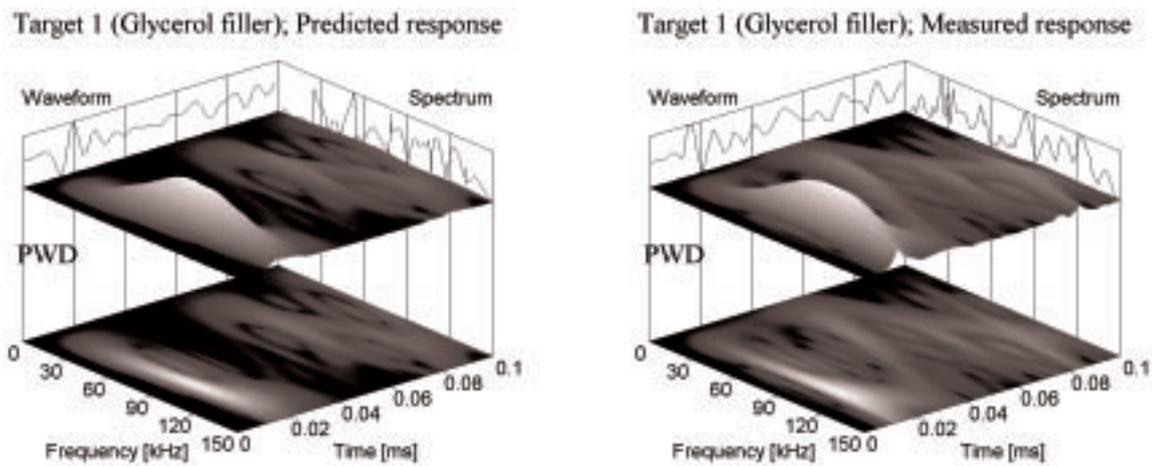

**Figure 1.** PWD of the theoretically predicted (left plot) or measured (right plot) echo from the shell filled with glycerol.

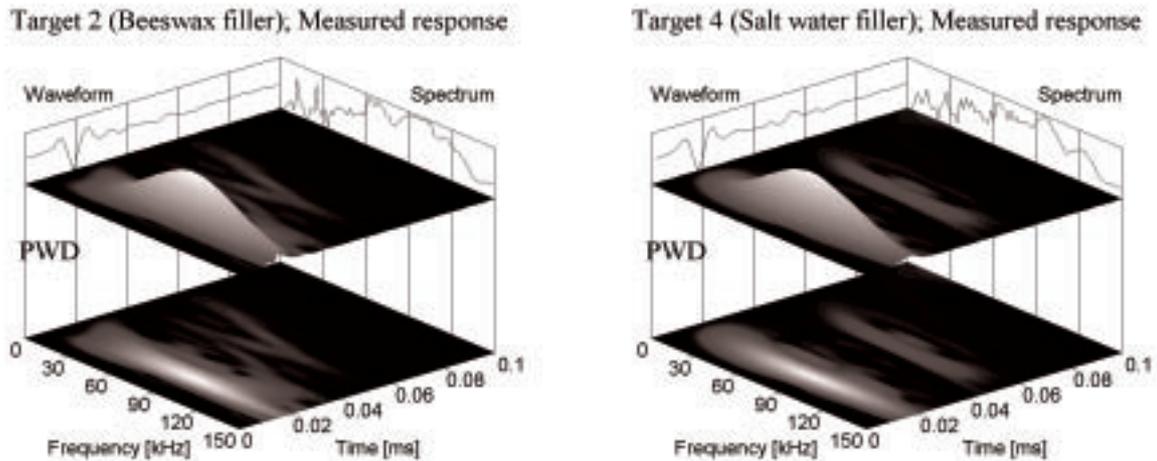

**Figure 2.** PWD of the measured echo from the shell filled with beeswax (left plot) or salt water (right plot).

## IV. APPLICATION TO MICROWAVES

### 1. GPR USES TO IDENTIFY BURIED LAND MINES

To analyze target responses in the joint time-frequency domain, we require the backscattered far-field $E_{sc}(t)$, or the spectrum of the illuminating pulse $G(\omega)$ together with the form-function $f_\infty(\omega)$, which are related by Eq. (2.7). These are the signals that are introduced in whichever distribution one wishes to use for the time-frequency analysis of the returned signals. As in the examples to be shown below, we will use the PWD with a suitable Gaussian window. The width of the window depends on the type of feature one wishes to emphasize. We have illustrated elsewhere [12] how the amount of signature features can be controlled using windowing technique. We will concentrate on scattering data obtained from objects buried underground and illuminated with short pulses of a (broadband) ground-penetrating radar (GPR) [13]. This is basically an ultra-wideband or impulse radar for land-mine detection and, particularly, classification.

### 2. INSTRUMENTATION AND ACCURACY OF MEASUREMENTS MADE WITH GPR

An impulse radar system was used to examine [14–17] ground penetration experiments when several targets were buried in the sand of an indoor large sand container of length / width / depth, 5 / 5 / 2 m. The sand container is at the Linköping site of the Swedish Defense Research Agency (FOI). The pulses transmitted by the impulse radar system were 0.3 ns in duration with a pulse-repetition frequency of 250 kHz and a peak power of 50 W. Two different antenna units were used both containing two broadband (200–2,000 MHz [14,15] or 200–3,000 MHz [16,17]) dipoles (transmitting and receiving), which are perpendicular. Data were sampled and pre-processed using a digitizing signal analyzer with a resolution of 14 bits. The equivalent time-sampling rate was chosen to be 20 GS/s (gigasamples per second). To use the system as a GPR the antenna unit was mounted on a positioning system above the ground surface. The antenna position was computer-controlled in three directions. The antenna was usually placed at a height of 2 or 5 cm above the ground surface. The strong specular return from the ground surface is suppressed using crossed dipoles, which also reduce echoes returned from many targets when the antennas are kept right above the target. Whatever coupling was present was suppressed by taking measurements with the antenna at a height of 70 cm above the ground (without any buried target) and then subtracting the recorded time-series from each one of the time-series obtained when a target was present. Ten sampled waveforms were ensemble averaged to suppress noise.

## 3. TIME-FREQUENCY SIGNATURES OF BURIED TARGETS

Given a simple target, it is of interest to examine its backscattered echoes using various time-frequency distributions in an attempt to determine which one offers the most advantages for classification. This is not an unambiguous task. In many instances two distributions can be equally satisfactory, and in other instances, this may not be the case. To illustrate this point we examine echoes from a metal sphere 10 cm in diameter buried in sand at a depth of either 5 cm or 15 cm. The test was conducted in the above-mentioned sandbox and with the center of the antennas located at a small horizontal (i.e., about 20 cm) distance from the vertical symmetry axis of the target. We have compared the echoes as plotted in the time-frequency domain by means of many distributions listed elsewhere [2]. These include the PWD and CWD given by Eqs. (2.3) and (2.4). Others were the "adaptive spectrogram" (AS), the "cone-shaped distribution" (CSD), the "Gabor spectrogram" (GS), and the "spectrogram" (SPEC). When applied to returned echoes from objects illuminated by an impulse radar the PWD, CWD and AS all have good resolution and features concentration, while simultaneously they all extract a reasonable amount of features (depending on the time and frequency resolution). These were found to best handle situations in which the target response is corrupted by clutter and multiple scattering effects, which are always present when underground targets are considered. The other distributions (i.e., CSD, GS and SPEC), all share the undesirable property of smearing out the extracted features [15]. The PWD seems to be robust enough to retain the main features of the target signature even when the burying depth is increased to 25 cm.

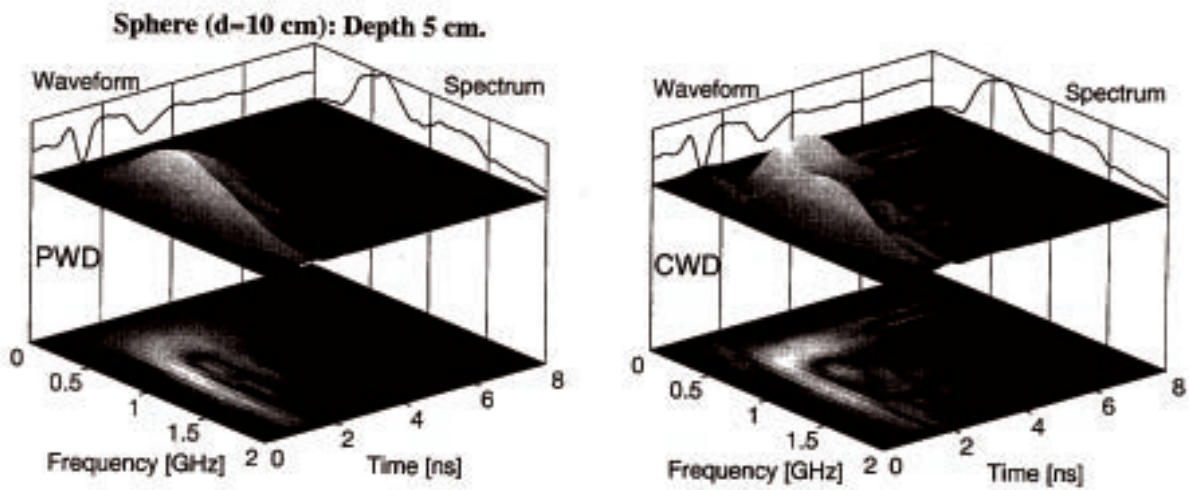

**Figure 3.** PWD (left plot) and CWD (right plot) of the backscattered echo from a 10 cm metal sphere buried in sand

at a depth of 5 cm.

Figure 3, left (or right) plot shows the PWD (or CWD) of the returned echo when a 10 cm metal sphere buried at the depth of 5 cm is illuminated by a GPR. The plots also show the waveform and its spectrum on the lateral planes. To appreciate the relative robustness of the PWD and CWD we compare Fig. 3 with the corresponding results when the burying depth is increased to 15 cm as shown in Fig. 4. Ideally, the time-frequency distributions in Fig. 4 should occur at a later time that corresponds to the larger burying depth while their shape would be retained. This is more or less the case, particularly when the echoes are analyzed using the PWD (left plots). While the characteristic shape of the CWD is retained, the CWD seems nonetheless to be more sensitive to the larger amount of clutter that is present at the 15 cm depth (right plots). The analyzed waveform is the same for both time-frequency distributions in Figs. 3 and 4, respectively. In view of its advantages we will use the PWD in the cases to follow in the next sections.

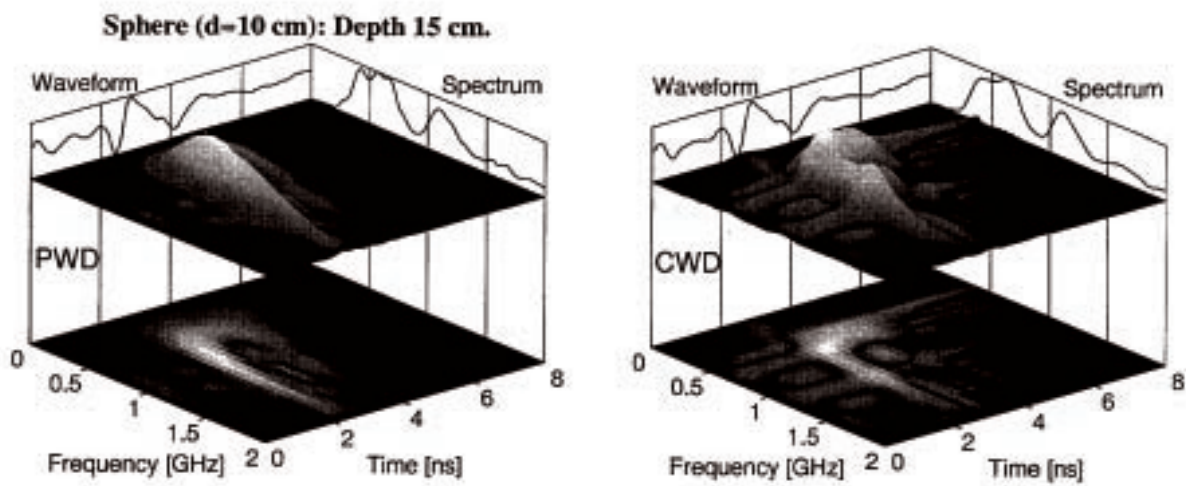

**Figure 4.** PWD (left plot) and CWD (right plot) of the backscattered echo from a 10 cm metal sphere buried in sand at a depth of 15 cm.

## 4. *FUZZY-CLUSTER REPRESENTATION OF TARGET SIGNATURES*

Any PWD calculated using a discrete-time version of Eq. (2.3) is then known in a rectangular grid defined by two integer coordinates [16,17]. Within each grid mesh the value of the PWD is constant and, if smaller than a selected value, can be set equal to zero to suppress the influence of signal noise or clutter. Each mesh with a nonzero value of the PWD is then represented by a point that is being randomly allocated within the mesh and associated with a weight equal to the value of the PWD. In this way we obtain a weighted point cluster representation of the PWD that can be used in conjunction with a fuzzy classification model based on cluster estimation [18]. The "fuzzy C-means" (FCM) clustering algorithm, modified to account for the weight associated with each cluster point, is then used to reduce the weighted point cluster representation of the PWD to a much smaller number of "cluster centers." The idea is that any set of points that are distributed over a surface of finite extent can be represented by much smaller number of cluster centers. In a "fuzzy" way these centers lie as close as possible to all the original given points. This could be achieved by first introducing a cost function that is defined to be the sum of each point's squared distance to the most adjacent cluster center, and then adjusting the position of each cluster center so as to minimize the cost function. All points that are closest to a given cluster center form a subcluster where these points have unit membership. If the cost function is modified by introducing a slightly different membership function that instead associates each point with a partial membership of each cluster center, then we have the full concept of the fuzzy C-means algorithm. The set of cluster centers computed using this algorithm is then a representation of the PWD that can be used as a template for target recognition purposes.

## 5. COMPARISON OF FUZZY-CLUSTER TEMPLATES WITH THE PWD

To illustrate the procedure [17], we consider five targets buried in the large sand container described above, with the GPR antenna unit with crossed dipoles (200–3,000 MHz) at a height of 5 cm above the ground surface. The targets are: (A) a dummy mine filled with beeswax, of diameter 8 cm and height 3.5 cm, (B) a dielectric nylon cylinder (diameter 25 cm and height 5 cm), (C) a metal sphere (diameter 10 cm), (D) a short metal cylinder (diameter 30 cm and height 5.5 cm), and (E) a stone the shape and size of a coconut. For each one of these targets, a total number of 400 measurements were performed. To suppress noise, the time-series are each composed of 10 ensemble averaged, or time-overlaid, sampled waveforms. The antenna center was positioned at each point of a grid centered above the target consisting of 20 lines in both directions with a 2 cm separation. The antenna elements made an angle of 45° with the grid lines. The PWD is evaluated from the discrete counterpart of Eq. (2.3) for each one of the five targets buried at 5 cm depth and illuminated by the GPR. The modified FCM algorithm is then used to generate a set of cluster centers.

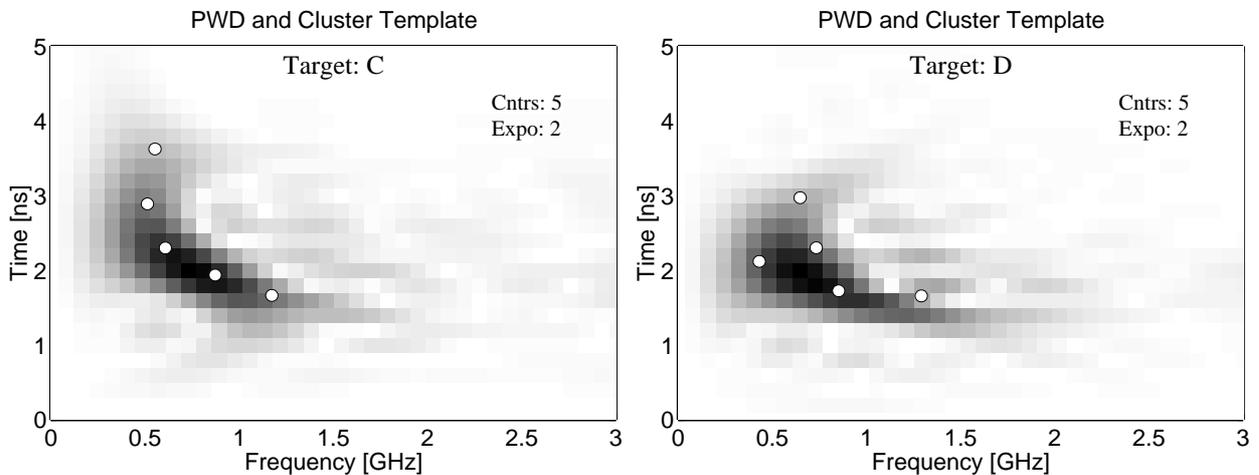

**Figure 5.** Comparison of the PWDs and the cluster templates for targets C and D.

Figure 5 displays surface plots of the PWDs for two of the targets (viz., C and D). The modulus of the (normalized) PWD is represented in shades of gray from 0 (white) to 1 (black). These plots show also the cluster centers, chosen to be five, that result from the application of the FCM algorithm modified to take weighted cluster points into account. These templates are now used for classification. For each recorded waveform returned from a target its PWD is computed.

Then each one of the templates is made to slide along the time axis in small (0.2 ns) steps using a time-window of width 5 ns. PWD values outside the time window are excluded to suppress clutter or signal contributions from multiple scattering. At each position of the template and time-window the weighted cost function is computed. The system keeps track of the smallest value and the template position where it occurred. This classifier does not utilize the strength of the returned echoes. Only the shape and relative magnitude of the generated PWD are used here.

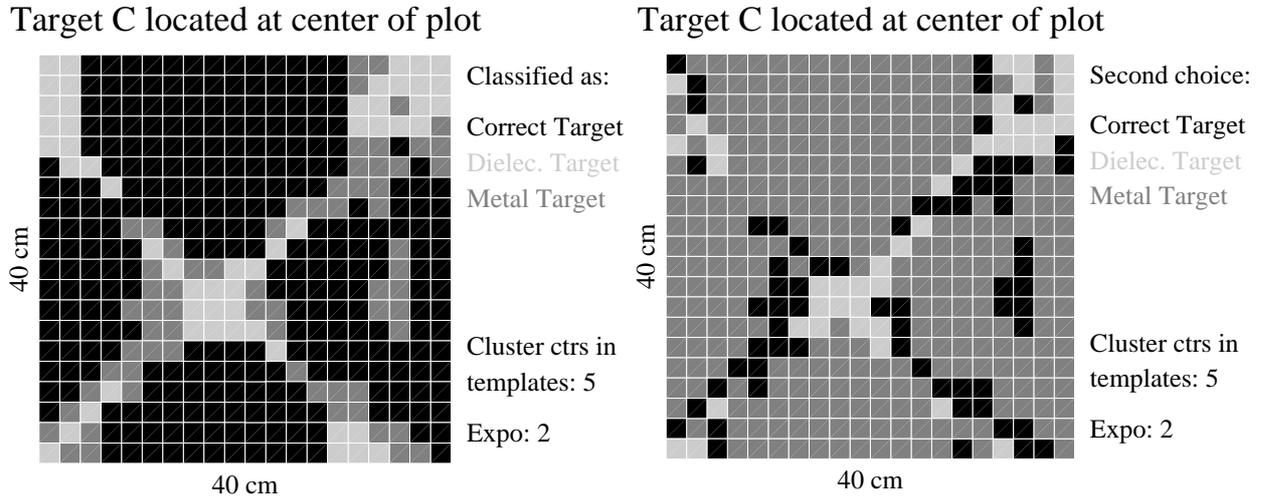

**Figure 6.** Classification results for target C (metal sphere).

Figure 6, left plot, displays the classification results of target C (i.e., the metal sphere), and we note that it is correctly classified at all antenna positions except along the two crossing diagonals. At these antenna positions, at least one of the antenna elements is reaching above the target's center and, because of the target's symmetry, the returned echo from the target is weak. The right plot of Fig. 6 displays the classifier's second choice. Where the target is correctly classified, the second choice is almost everywhere the other metal target (i.e., D, the short metal cylinder). Similarly, Fig. 7 displays the classification results for target D. In this case the target is more often misclassified as the other metal target (i.e., C). Hardly anywhere except close to the crossing diagonals is any of these targets misclassified as one of the dielectric targets. Conversely, the dielectric targets (i.e., A, B, and E) were most often classified as another dielectric target when misclassified [17] (not shown here).

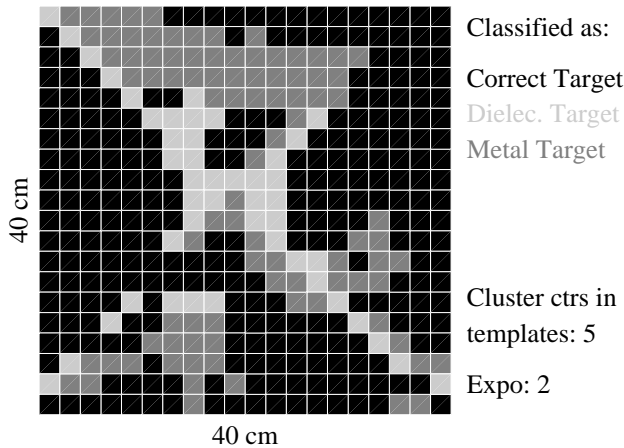 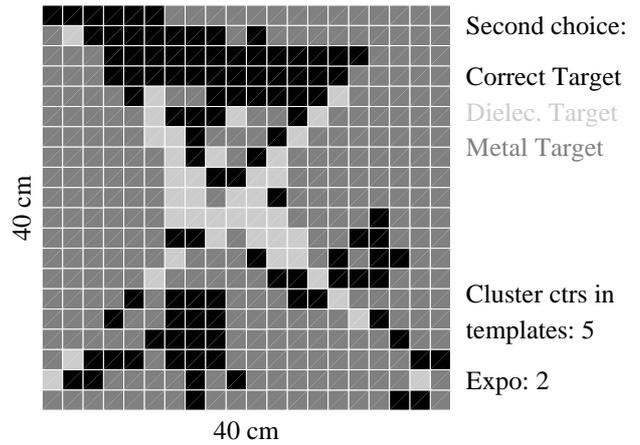

**Figure 7.** Classification results for target D (short metal cylinder).

## V. SUMMARY AND CONCLUSIONS

We have shown how impulse sonars or radars are, or can be, used for target identification purposes. We illustrated several useful approaches in the joint time-frequency domain. The identifying resonance features are best analyzed in the time-frequency domain using bilinear distributions of the Wigner-type. A favorite one in this work was the pseudo-Wigner distribution obtained by introducing a smooth time-window in the Wigner distribution. The extracted features have been connected in many instances to actual physical characteristics of the targets. In the present review we showed a sonar example in which we proposed and explained how these techniques could elucidate the amazing classification feats performed by dolphins. We also discussed the operation of an impulse radar functioning as a ground penetrating radar to identify subsurface land mines.

## ACKNOWLEDGEMENTS

The authors gratefully acknowledge the partial support of their respective Institutions. This work was partially supported by the Office of Naval Research and the Swedish Rescue Service Agency.